\documentclass[12pt,twoside]{article}
\usepackage{xrb2007}
\usepackage{cite}
\usepackage{epsfig}
\usepackage{graphicx}
\usepackage{amsmath}
\begin{document}

\session{Jets}

\shortauthor{Miller-Jones et al.}
\shorttitle{Jet-ISM interactions in X-ray binaries}

\title{Searching for the signatures of jet-ISM interactions in X-ray binaries}
\author{James Miller-Jones}
\affil{Jansky Fellow, National Radio Astronomy Observatory, 520 Edgemont Road, Charlottesville, VA
  22902, USA}
\author{Christian Kaiser, Tom Maccarone, Rob Fender, Anna Kapi\'nska,
  Katherine Gunn}
\affil{School of Physics and Astronomy, University of Southampton,
  Highfield, Southampton SO17 1BJ, UK}
\author{Dave Russell}
\affil{Astronomical Institute `Anton Pannekoek', University of
  Amsterdam, Kruislaan 403, 1098 SJ, Amsterdam, The Netherlands}
\author{Catherine Brocksopp}
\affil{MSSL, University College London, Holmbury St.\ Mary, Dorking,
  Surrey RH5 6NT, UK}
\author{Jennifer Sokoloski}
\affil{Columbia Astrophysics Lab, MC 5247, Columbia University, New
  York, NY 10027, USA}
\author{Ben Stappers, Tom Muxlow}
\affil{University of Manchester, Jodrell Bank Observatory,
  Macclesfield, Cheshire SK11 9DL, UK}

\begin{abstract}
Jets from X-ray binaries are continuously injecting matter and energy
into the surrounding interstellar medium.  However, there exist to
date relatively few cases where jet-ISM interactions have been
directly observed.  We review the current examples, and go on to
present new data on the proposed hotspots of GRS\,1915+105, finding no
concrete evidence for any association between the hotspots and the
central source, in agreement with previous findings in the literature.
We also present preliminary results on radio and H$\alpha$ searches
for jet-ISM interactions around known X-ray binaries, and discuss
strategies for future searches.
\end{abstract}

\section{Introduction}
The jets in X-ray binary systems have been estimated to inject
$\sim1$\% of the time-averaged luminosity of supernovae into the
surrounding interstellar medium (ISM) \citep{Fen05_jmj}.  Heinz
\citetext{these proceedings} has estimated that the jets also inflate
lobes at a rate of $\sim5\times10^{48}$\,cm$^{3}$\,s$^{-1}$,
comparable to a significant fraction of the volume of the Galactic
disc over the lifetime of the Galaxy.  Furthermore, the $\mu$G-level
magnetic fields they inject into their surroundings could be
responsible for seeding the Galactic magnetic field.  However, despite
the undisputed importance of the effects of X-ray binary jets on their
environments, there are as yet relatively few cases where the
interaction between the jets and the surrounding ISM has been directly
observed.

Extended, stationary radio lobes have been observed around the two
systems 1E\,1740.7-2942 \citep{Mir92_jmj} and GRS\,1758-258
\citep{Mar02_jmj}.  In three further cases, the large-scale nebulae
inflated by the jets have been imaged in the radio band.  The
precessing jets of SS\,433 are thought to have inflated the two `ears'
of the surrounding W\,50 nebula \citep{Beg80_jmj}, a radio lobe
aligned with the jets of Cyg X-1 was recently discovered by
\citet{Gal05_jmj}, and in Cir X-1, we observe the radio jets through
the nebula which they have inflated \citep{Tud06_jmj}.
\citet{Fom01_jmj} monitored the radio lobes of Sco X-1 as they moved
outwards, which were identified as the working surfaces where
highly-relativistic beams impacted on the ambient medium.  A similar
case of a relativistic underlying flow lighting up downstream radio
lobes has been observed in a second confirmed neutron-star system, Cir
X-1 \citep{Fen04_jmj}, and it has been suggested that a similar unseen
shock is responsible for lighting up the X-ray jets in SS\,433
\citep{Mig05_jmj}.  Such impact sites have been directly detected in
the X-ray band in the cases of XTE J\,1550-564 \citep{Cor02_jmj} and
H\,1743-322 \citep{Cor05_jmj}, at angular separations of several
arcseconds from the respective central binary systems.  Actual
deceleration of the jets as they sweep up the surrounding ISM has only
been unambiguously observed in XTE J\,1550-564 \citep{Kaa03_jmj},
although evidence for deceleration has also been seen in the radio
band in the system XTE J\,1748-288.
\begin{figure}
\plotone{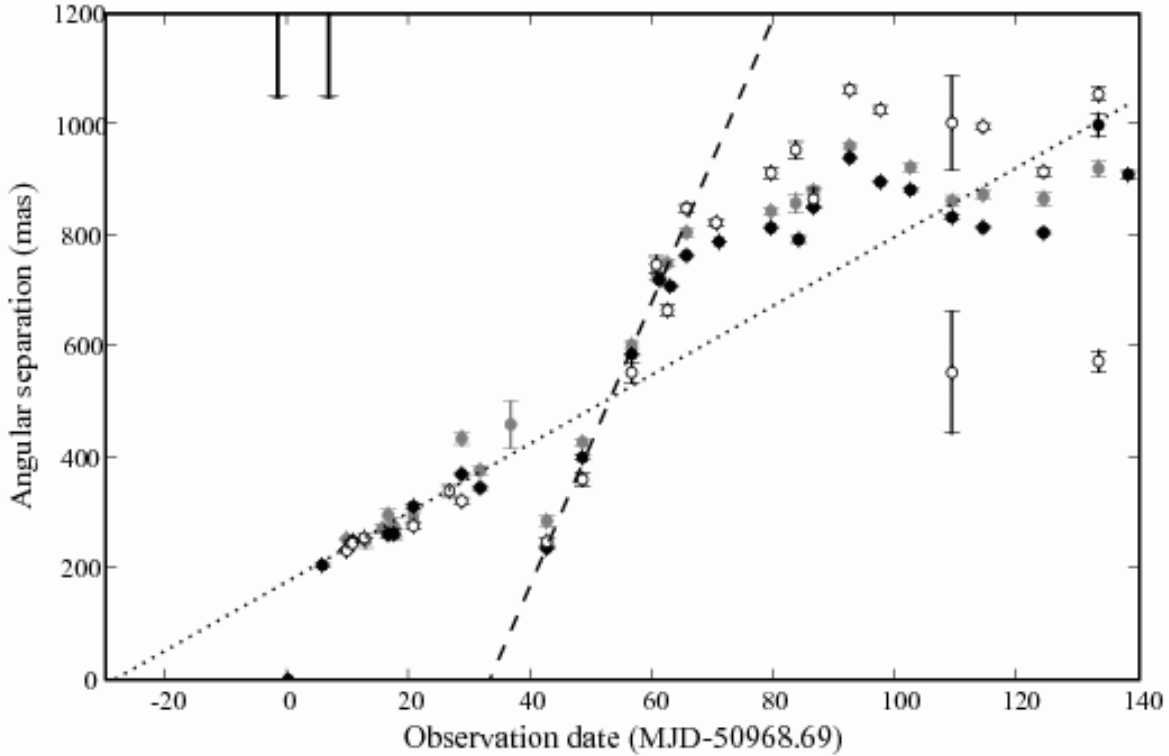}
\caption{Measured angular separation from the core of the radio knots
  in the 1998 outburst of XTE J\,1748-288.  Radio observations were
  made at 5, 8.4 and 15\,GHz, and day 0 is the ejection date inferred
  from X-ray observations.  Open symbols are for 15\,GHz, filled black
  symbols for 8.4\,GHz, and filled grey symbols are for 5\,GHz.  The
  two arrows indicate the start and peak of the X-ray outburst
  \citep{Bro07_jmj}.}
\label{fig:xte1748}
\end{figure}

Figure \ref{fig:xte1748} shows the measured angular separations of the
radio knots from the core, as a function of time since the start of
the corresponding X-ray flare during the 1998 outburst of XTE
J\,1748-288.  There must have been deceleration within 200\,mas,
before the first radio observations were made, since the fitted radio
proper motion of 6.2\,mas\,d$^{-1}$ does not give a zero-separation
date consistent with the X-ray flare.  The proper motion then
increases to 25.5\,mas\,d$^{-1}$ for $\sim20$\,d before the jets seem
to stall once more at an angular separation of $\sim1$\,arcsec.  This
is a clear indication of complex interactions with the environment,
and further analysis is warranted.  Higher-frequency observations
suggest multiple ejection events, which could help to explain the
observed speeds, although the system is located in the direction of
the Galactic Centre, so if it is not a foreground source, the
surrounding environment is likely to be both dense and highly
inhomogeneous, which could also explain the variable proper motions.

\section{The proposed hotspots of GRS\,1915+105}
\label{sec:grs1915}
GRS\,1915+105 is the prototypical microquasar, and has been in a
continuous state of outburst since 1992.  During its so-called plateau
state, it exhibits a flat-spectrum conical jet with luminosity
$\sim10^{38}$\,erg\,s$^{-1}$, but it periodically undergoes outbursts
when relativistically-moving ejecta are observed and the jet power
rises to $\sim10^{41}$\,erg\,s$^{-1}$ \citep{Fen04a_jmj}.  Despite
being one of the most powerful relativistic jets known, no unequivocal
evidence for interactions with the ISM has been seen in this system,
possibly owing to its location in a relatively underdense region
\citep{Hei02_jmj}.  A discrepancy in the proper motions of the jets
measured on different angular scales \citep{Mir94_jmj,Fen99_jmj} was
initially suggestive of deceleration within a few hundred
milliarcseconds of the central binary, but \citet{Mil07_jmj},
collating all available data on the relativistic outbursts of the
source, found no conclusive evidence for deceleration.

\citet{Rod98_jmj} examined the surroundings of GRS\,1915+105, and
tentatively identified two possible hotspots well-aligned with the
arcsecond-scale jets and equidistant from the source at an angular
separation of 17\,arcmin.  \citet{Cha01_jmj} observed these sources in
the radio, millimeter and infrared bands and found no clear evidence
that they were associated with GRS\,1915+105.  More recently,
\citet{Kai04_jmj} presented a self-consistent model of these two
sources as hotspots, identifying the non-thermal filament pointing
back from the southwestern hotspot to the central binary as emission
from the end of the jet where it impinged on the ambient medium.
Taking the perspective that absence of evidence is not in itself
evidence of absence, we made further follow-up observations of the
south-western hotspot, IRAS 19132+1035, to test this hypothesis using
the VLA and MERLIN arrays and the {\it Chandra} X-ray telescope.

\begin{figure}
\includegraphics[width=\columnwidth]{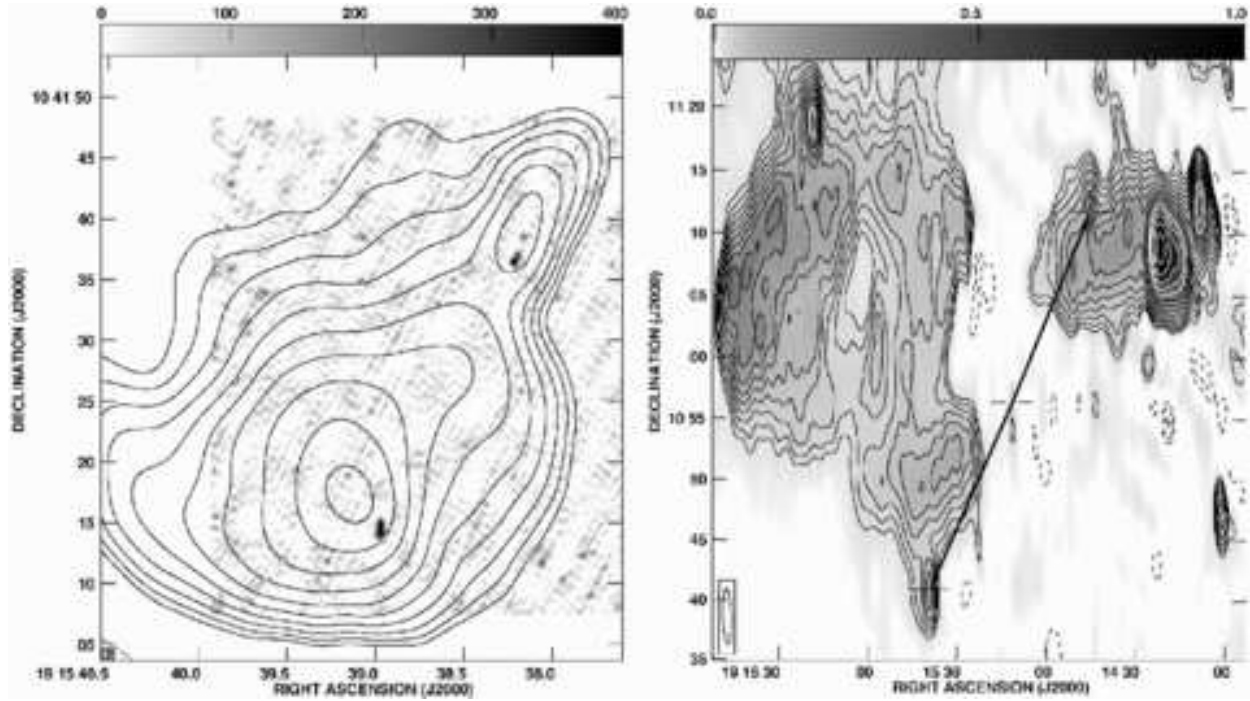}
\caption{{\it Left}: Overlay of the VLA 1.4-GHz contours on the MERLIN
  greyscale image of the southeastern hotspot candidate source
  IRAS\,19132+1035.  Contours are at levels of $\pm(\sqrt{2})^n$ times
  300\,$\mu$Jy\,bm$^{-1}$, where $n=3,4,5,...$, and the greyscale is
  in $\mu$Jy\,bm$^{-1}$.  {\it Right}: WSRT 350-MHz image of the field
  surrounding GRS\,1915+105, with crosses marking the positions of the
  central source and the candidate hotspots.  The greyscale is in
  Jy\,bm$^{-1}$ and the contours are at levels of $\pm(\sqrt{2})^n$
  times the rms of 3.9\,mJy\,bm$^{-1}$, where $n=3,4,5,...$}
\label{fig:grs1915}
\end{figure}

VLA 1.4\,GHz observations showed that the flux density of the proposed
hotspot had not changed between 1997 and 2005, despite at least 9
flaring sequences occurring in the central source during this time.
However, given the propagation time of $\geq25$\,y for jet material to
travel down the jet to the candidate hotspot, then since the source
only switched on in 1992, we might not necessarily expect to see any
variability of the hotspot region.  Alternatively, the observed
variability of the central source could be smoothed out during its
propagation down the jet.  To explore the nature of the non-thermal
filament, and determine whether it could simply be a background AGN
coincidentally located close to the hotspot and oriented along the
position angle of the arcsecond-scale jets of GRS\,1915+105, we made
high-resolution radio observations with MERLIN.  Only 20\% of the
1.4-GHz radio flux density seen with the VLA was recovered, suggesting
that the majority of the emission from the non-thermal filament was
from a more diffuse, large-scale structure which was being spatially
filtered out on the longer MERLIN baselines.  This would appear to be
inconsistent with an AGN interpretation.  Also, the morphology of the
observed emission does not resemble the typical core-hotspot structure
of a background FR\,II radio galaxy (Fig.~\ref{fig:grs1915}, left
panel).  No spatially-extended bowshock structure was observed from
the southern border of the VLA source, although the jet model of
\citet{Kai04_jmj} makes no prediction as to the morphology of the
hotspot region.  The {\it Chandra} observations detected no
significant emission at the position of the candidate hotspot ($8\pm8$
counts; consistent with zero, but also at the $3\sigma$ level with
half the 60 counts expected from an extrapolation of the radio
measurement, assuming a standard synchrotron spectrum), suggesting
that if this is the interaction region, any shock acceleration is
weaker than has been seen in sources such as XTE J\,1550-564, and is
insufficient to power X-ray lobes.  Thus even with the new data, we
still cannot either fully confirm or deny an association between the
candidate lobes and the central binary, just as found by \citet{Cha01_jmj}.
It should be noted that the H\,{\textsc i} distance to the two hotspot
candidates is 6.5\,kpc, significantly closer than the $\sim11$\,kpc
that is usually assumed for GRS\,1915+105.  Recent work by
\citet{Dha07_jmj} found that the peculiar motion of the central binary
system is minimized for a source distance of 9\,kpc, which would rule
out any association with the candidate hotspots, although the authors
do not exclude a closer distance should the system have received a
substantial natal kick during the formation of the black hole.

\section{Searching for lobes with low-frequency radio observations}
In order to look for evidence of jet-blown lobes surrounding three of
the most powerful known X-ray binary systems, and to characterise
their behaviour at low frequencies, we observed the fields surrounding
GRS1915+105, Cyg X-3, and SS\,433 with the WSRT at 350 and 140\,MHz.
The right-hand panel of Fig.~\ref{fig:grs1915} shows the field
surrounding GRS\,1915+105.  While the two hotspot candidates described
in Section 2 were both detected, with spectra consistent with those
found at higher frequencies by \citet{Cha01_jmj}, there was no
evidence for any extended emission linking them back to the central
binary system.  This is consistent with the findings of
\citet{Kai04_jmj}, who predicted that the synchrotron luminosity of
any such lobes ought to be of order 0.08\,mJy\,bm$^{-1}$ at 1.4\,GHz,
below the sensitivity of the current generation of radio telescopes.

\begin{figure}
\includegraphics[width=\columnwidth]{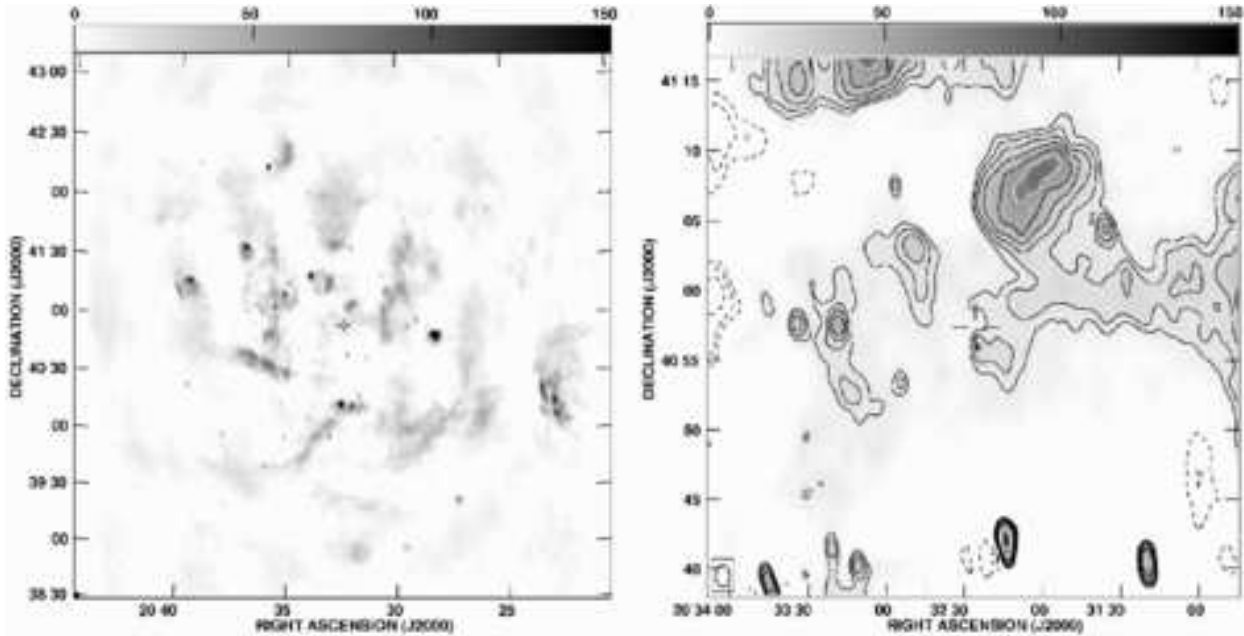}
\caption{{\it Left}: WSRT-350-MHz image of the field surrounding
  Cygnus X-3.  Greyscale is in mJy\,bm$^{-1}$.  {\it Right}: Zoomed-in
  image of the region surrounding the binary system.  The greyscale is
  in mJy\,bm$^{-1}$ and the contours are at levels of
  $\pm(\sqrt{2})^n$ times the rms of 2.2\,mJy\,bm$^{-1}$, where
  $n=3,4,5,...$.  In both panels, the cross marks the location of the
  central binary system.}
\label{fig:cygx3}
\end{figure}

In Cygnus X-3, there was also no evidence for any extended lobes.  The
line of sight passes through the Cygnus OB2 association, a dense field
with significant extended emission and many point sources.  Figure
\ref{fig:cygx3} shows the field surrounding the X-ray binary, with the
full WSRT 350-MHz field of view on the left and a zoomed-in version on
the right.  It is clear that there is too much diffuse emission in the
field to unambiguously associate any structures with the central
binary system.  The central binary was not detected at either
frequency, and a comparison to 15-GHz monitoring data from the Ryle
Telescope (Pooley; \texttt{http://www.mrao.cam.ac.uk/$\sim$guy/cx3/})
implies that there must be a turnover in the spectrum, either due to
self-absorption or free-free absorption.  The source was however
significantly detected in later observations during its giant radio
outburst of 2006 at a level of 2\,Jy at 140\,MHz.

\section{Searching for lobes with H$\alpha$ observations}
The jet-blown lobe in Cyg X-1 was detected both in the radio band and
in deep narrowband optical images taken in the H$\alpha$ and
[O\,\textsc{iii}] (5007\,\AA) filters \citep{Rus07_jmj}.  The
luminosity and morphology of the latter emission is indicative of
shock-excited gas with velocity $v_{\rm shock} > 100$\,km\,s$^{-1}$.
A search for optical nebulae around X-ray binaries \citep{Rus06_jmj}
identified three promising candidates around the sources GRO
J\,1655-40, GRS\,1009-45, and LS\,5039.  A cavity in H$\alpha$ around
GRO J\,1655-40 was observed, and a shell which aligned well with
diffuse radio emission in the region.  The lobe was also detected in
[S\,\textsc{ii}] emission, a tracer of shock-excited gas, and
low-frequency 843-MHz MOST emission appeared to trace the outline of
the cavity well.  Deeper VLA observations of GRO J\,1655-40 and
LS\,5039 have been taken in order to attempt to unambiguously
determine whether the cavities are real, and whether they are in fact
associated with their respective central binaries.  In one further
(extragalactic) case, the H$\alpha$ nebula surrounding LMC X-1 is
likely to be both photoionized by the X-ray source and shock-excited
by its jets \citep{Coo07_jmj}.

\section{Observing strategies}
Extensive observation campaigns have thus far failed to identify many
examples of jet-ISM interactions or jet-blown lobes around X-ray
binary systems.  There is still no evidence for the effects of some of
the most powerful known jets, such as those of GRS\,1915+105, on their
surroundings.  Since the time-averaged power output of such transient
jets is thought to be comparable to that of the compact, steady jets
\citep{Hei05_jmj}, there is no reason to target only the brightest
systems in a search for such interactions.  As argued by
\citet{Hei02_jmj}, since most X-ray binaries are located in
dynamically underdense environments when compared to AGN, local
density enhancements are required to slow the jets via
interaction with the ISM.  A further possibility is that in order to
inflate jet-blown lobes, the central source must have a low peculiar
velocity relative to the local standard of rest.  Cyg X-1 has a
velocity of only $9\pm2$\,km\,s$^{-1}$ relative to the nearby Cygnus
OB\,3 association \citep{Mir03_jmj}, such that the effectively
constant jet direction has allowed it to inflate the large-scale lobe over
its 0.02--0.06\,Myr lifetime \citep{Rus07_jmj}.  For other X-ray
binaries which received higher natal kicks during their formation, the
jet direction might not be sufficiently stable to inflate large-scale
nebulae, although this would not rule out interaction with the ISM at
the points of jet impact, as seen in XTE J\,1550-564 or Sco X-1.

In order to detect extended emission surrounding X-ray binaries in the
radio band, it is necessary to observe at low frequencies in compact
configurations to probe the diffuse emission from the jet-blown
nebulae.  Synchrotron emission will be brightest at lower frequencies,
and deep integrations are needed to pick out low surface brightness
emission.  Detecting polarized emission from such lobes could help to
prove that the emission was of synchrotron origin from the high-energy
electrons in the jets, filtering out unwanted thermal emission from
the complex Galactic fields in which many X-ray binaries are located.
Instruments such as LOFAR and the MWA, with their wide fields of view,
could help in detecting more candidate objects.  However, extended
radio emission alone cannot necessarily prove an association between
any candidate nebulae and their central objects, and a multiwavelength
approach, such as combining radio with narrowband optical
observations, is often required to definitively identify jet-inflated
lobes.

\section{Conclusions}
X-ray binary jets are continuously injecting energy, momentum and
magnetic fields into their surroundings, and the signs of these
interactions can be directly observed in some cases.  However,
detection of such interactions requires fairly special conditions, in
particular low peculiar velocities and local density enhancements in
the environments of the sources.  A second, in-depth, multiwavelength
study has revealed no firm evidence for jet-blown lobes around
GRS\,1915+105, and a secure distance determination (e.g.\ via
trigonometric parallax) appears to be the only way to conclusively
confirm or rule out an association with the two IRAS sources
identified by \citet{Rod98_jmj}.

\acknowledgments Support for this work has been provided in part by
NASA through Chandra Award SAO G06-7026X-R.  JM-J is a Jansky Fellow
of the NRAO.  The NRAO is a facility of the NSF operated under
cooperative agreement by Associated Universities, Inc.


\vspace{-0.2cm}
\begin{thebibliography}{}

\bibitem[Begelman et al.(1980)]{Beg80_jmj}
Begelman M.~C., Hatchett S.~P., McKee C.~F., Sarazin C.~L., Arons J., 1980, ApJ, 238, 722

\bibitem[Brocksopp et al.(2007)]{Bro07_jmj}
Brocksopp C., Miller-Jones J.~C.~A., Fender R.~P., Stappers B.~W., 2007, MNRAS, 378, 1111

\bibitem[Chaty et al.(2001)]{Cha01_jmj}
Chaty S., Rodr\'\i guez, L.~F., Mirabel I.~F., Geballe T.~R., Fuchs Y., Claret A., Cesarsky C.~J., Cesarsky D., 2001, A\&A, 366, 1035

\bibitem[Cooke et al.(2007)]{Coo07_jmj}
Cooke R., Kuncic Z., Sharp R., Bland-Hawthorn J., 2007, ApJ, 667, L163

\bibitem[Corbel et al.(2002)]{Cor02_jmj}
Corbel S., Fender R., Tzioumis A., Tomsick J., Orosz J., Miller J., Wijnands R., Kaaret P., 2002, Science, 298, 196

\bibitem[Corbel et al.(2005)]{Cor05_jmj}
Corbel S., Kaaret P., Fender R., Tzioumis A., Tomsick J., Orosz J., 2005, ApJ, 632, 504

\bibitem[Dhawan et al.(2007)]{Dha07_jmj}
Dhawan V., Mirabel I.~F., Rib\'o  M., Rodrigues I., 2007, ApJ, 668, 430

\bibitem[Fender \& Belloni(2004)]{Fen04a_jmj}
Fender R., Belloni T., 2004, ARA\&A, 42, 317

\bibitem[Fender et al.(1999)]{Fen99_jmj}
Fender R.~P., Garrington S.~T., McKay D.~J., Muxlow T.~W.~B., Pooley G.~G., Spencer R.~E., Stirling A.~M., Waltman E.~B., 1999, MNRAS, 304, 865

\bibitem[Fender et al.(2004)]{Fen04_jmj}
Fender R., Wu K., Johnston H., Tzioumis T., Jonker P., Spencer R., van der Klis M., 2004, Nat, 427, 222

\bibitem[Fender et al.(2005)]{Fen05_jmj}
Fender R.~P., Maccarone T.~J., van Kesteren Z., 2005, MNRAS, 360, 1085

\bibitem[Fomalont et al.(2001)]{Fom01_jmj}
Fomalont E.~B., Geldzahler B.~J., Bradshaw C.~F., 2001, ApJ, 553, L27

\bibitem[Gallo et al.(2005)]{Gal05_jmj}
Gallo E., Fender R., Kaiser C., Russell D., Morganti R., Oosterloo T., Heinz S., 2005, Nat, 436, 819

\bibitem[Heinz(2002)]{Hei02_jmj}
Heinz S., 2002, A\&A, 388, L40

\bibitem[Heinz \& Grimm(2005)]{Hei05_jmj}
Heinz S., Grimm H.~J., 2005, ApJ, 633, 384

\bibitem[Kaaret et al.(2003)]{Kaa03_jmj}
Kaaret P., Corbel S., Tomsick J.~A., Fender R., Miller J.~M., Orosz J.~A., Tzioumis A.~K., Wijnands R., 2003, ApJ, 582, 945

\bibitem[Kaiser et al.(2004)]{Kai04_jmj}
Kaiser C.~R., Gunn K.~F., Brocksopp C., Sokoloski J.~L., 2004, ApJ, 612, 332

\bibitem[Mart\'\i\ et al.(2002)]{Mar02_jmj}
Mart\'\i, J., Mirabel I.~F., Rodr\'\i guez, L.~F., Smith I.~A., 2002, A\&A, 386, 571

\bibitem[Migliari et al.(2005)]{Mig05_jmj}
Migliari S., Fender R.~P., Blundell K.~M., M\'endez, M., van der Klis M., 2005, MNRAS, 358, 860

\bibitem[Miller-Jones et al.(2007)]{Mil07_jmj}
Miller-Jones J.~C.~A., Rupen M.~P., Fender R.~P., Rushton A., Pooley G.~G., Spencer R.~E., 2007, MNRAS, 375, 1087

\bibitem[Mirabel \& Rodr\'\i guez(1994)]{Mir94_jmj}
Mirabel I.~F., Rodr\'\i guez L.~F., 1994, Nat, 371, 46

\bibitem[Mirabel \& Rodrigues(2003)]{Mir03_jmj}
Mirabel I.~F., Rodrigues I., 2003, Science, 300, 1119

\bibitem[Mirabel et al.(1992)]{Mir92_jmj}
Mirabel I.~F., Rodriguez L.~F., Cordier B., Paul J., Lebrun F., 1992, Nat, 358, 215

\bibitem[Rodr\'\i guez \& Mirabel(1998)]{Rod98_jmj}
Rodr\'\i guez L.~F., Mirabel I.~F., 1998, A\&A, 340, L47

\bibitem[Russell et al.(2006)]{Rus06_jmj}
Russell D., Fender R., Gallo E., Miller-Jones J.~C.~A., Kaiser C.~R.,
2006, in ``VI Microquasar Workshop: Microquasars and Beyond'',
ed. T. Belloni, PoS (MQW6), 59

\bibitem[Russell et al.(2007)]{Rus07_jmj}
Russell D.~M., Fender R.~P., Gallo E., Kaiser C.~R., 2007, MNRAS, 376, 1341

\bibitem[Tudose et al.(2006)]{Tud06_jmj}
Tudose V., Fender R.~P., Kaiser C.~R., Tzioumis A.~K., van der Klis M., Spencer R.~E., 2006, MNRAS, 372, 417

\end{thebibliography}
\end{document}